\documentclass{PoS}
\usepackage{amsmath}
\usepackage{amssymb}
\usepackage{latexsym}
\usepackage{graphics}

\def\be{\begin{equation}}
\def\ee{\end{equation}}
\def\bea{\begin{eqnarray}}
\def\eea{\end{eqnarray}}
\def\ba{\begin{array}}
\def\ea{\end{array}}
\def\({\left(}
\def\){\right)}
\def\e{{\rm e}}
\def\tr{{\rm tr}}
\def\qdet{{\rm qdet}}

\def\nn{\nonumber}
\newcommand\bs[1]{\pmb{#1}}

\title{Distribution of the k$^{\rm th}$ smallest Dirac operator eigenvalue : an update}

\ShortTitle{Distribution of the k$^{\rm th}$ smallest Dirac operator eigenvalue : an update}

\author{\speaker{Shinsuke M. Nishigaki}\thanks{
This work is supported in part by JSPS Grants-in-Aid for Scientific Research (KAKENHI) No.\ 25400259. 
}\\
       Graduate School of Science and Engineering\\
       Shimane University, Matsue 690-8504, Japan\\
       E-mail: \email{mochizuki@riko.shimane-u.ac.jp}}

\abstract{
Based on the exact relationship to random matrix theory, we present an alternative method of 
evaluating the probability distribution of the $k^{\rm th}$ smallest Dirac eigenvalue
in the $\varepsilon$-regime of QCD and QCD-like theories.
By utilizing the Nystr\"{o}m-type discretization of Fredholm determinants and Pfaffians, 
practical trouble of evaluating multiple integrations is circumvented and 
technical restrictions on the parities of the number of flavors and of the topological charge
present in our previous treatment for $\beta=1$ and 4 cases
[Phys.\ Rev.\ D{\bf 63}, 045012 (2001)] are partly lifted.
This method is also applied to the distributions of spacings between
$k^{\rm th}$ nearest-neighboring levels in the mobility edges of Anderson Hamiltonian and Dirac operator
in high-temperature QCD.}

\FullConference{The 33rd International Symposium on Lattice Field Theory\\
		14 -18 July 2015\\
		Kobe International Conference Center, Kobe, Japan*}

\begin{document}

\section{Introduction}
Random matrix theory opened
a nonperturbative and analytical window
into the physics of
chiral symmetry breaking in QCD \cite{SV93}.
Namely, 
chiral Gaussian ensemble of Hermitian random matrices of the 
block off-diagonal form
$
H=
\(
{
\ 0 \  \ W
\atop
W^\dagger\; 0
}
\)
$
with 
$W\in \mathbb{R}^{N\times (N+\nu)}$, 
$\mathbb{C}^{N\times (N+\nu)}$, 
or
$\mathbb{H}^{N\times (N+\nu)}$
(labeled by the Dyson indices $\beta=1, 2, 4$, respectively),
that are distributed according to the unnormalized probability measure
\be
dW\,
\e^{-\beta \tr\, W^\dagger W}
\prod_{f=1}^n \mathrm{det}\(\begin{matrix}
m_f & i W\\
i W^\dagger & m_f \end{matrix}\),
\label{jpdchiralmassiv}
\ee
($\det\to \qdet$ for $\beta=4$)
exhibits
the spontaneous symmetry breaking pattern
expected for confining gauge theories coupled to
$n$ flavors of Dirac ($\beta=1, 2$) or Majorana ($\beta=4$) 
fermions in a pseudoreal, complex, or real representation,
in a sector of topological charge $\nu$ \cite{Ver94}.
The joint distribution of the square of $N$ positive eigenvalues of $H$
(singular values of $W$) $\{x_i\}$ then reads
\be
P(\{x\})=
\mathcal{Z}_{\nu}(\{m\})^{-1}
\prod_{i=1}^N
\Bigl(
x_i^{\beta(\nu+1)/2-1}\e^{-\beta x_i}
\prod_{f=1}^{n} (x_i+m_f^2)
\Bigr)
\prod_{j>k}^N|x_j-x_k|^\beta.
\label{chGbE}
\ee
The universal $k$-level correlation functions $R_k$
(including the spectral density $\rho(\zeta)$ for $k=1$)
of these `massive' ensembles,
expressed as determinantal/Pfaffian point processes \cite{DN98,NN00},
are not the quantities
that most suit to fit the Dirac spectral data from lattice QCD simulations,
as the amplitude of oscillation of $R_k$
quickly decreases
for a large separation between eigenvalues or from the origin.
Instead, the clear winner is the individual distributions $p_k(\zeta)$ 
of each ordered eigenvalues (Fig.~1), which are also universal \cite{NDW98}.
Each of them has a characteristic, quasi-Gaussian peaky form
that resolves the spectral density (the latter is almost structureless for $\zeta\gg1$)
as $\rho(\zeta)=\sum_{k\geq 1}p_k(\zeta)$.
\begin{figure}[b]
\begin{center}
\includegraphics[bb=0 0 342 198,width=82mm]{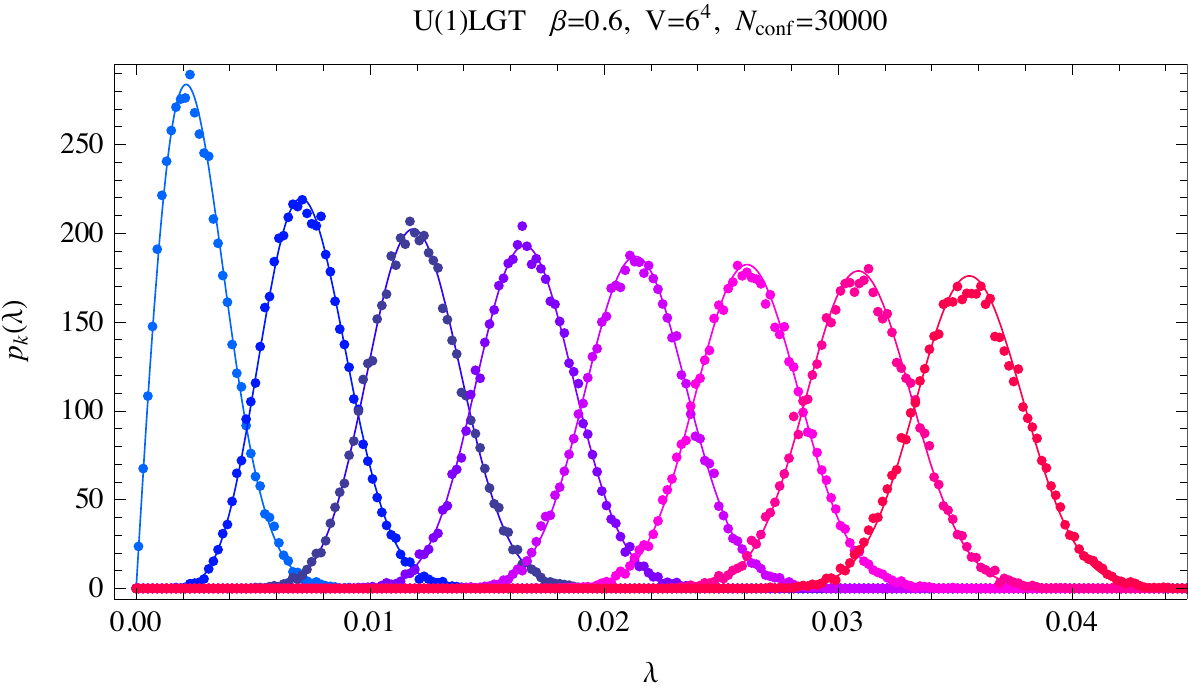}~
\includegraphics[bb=0 0 262 150,width=68mm]{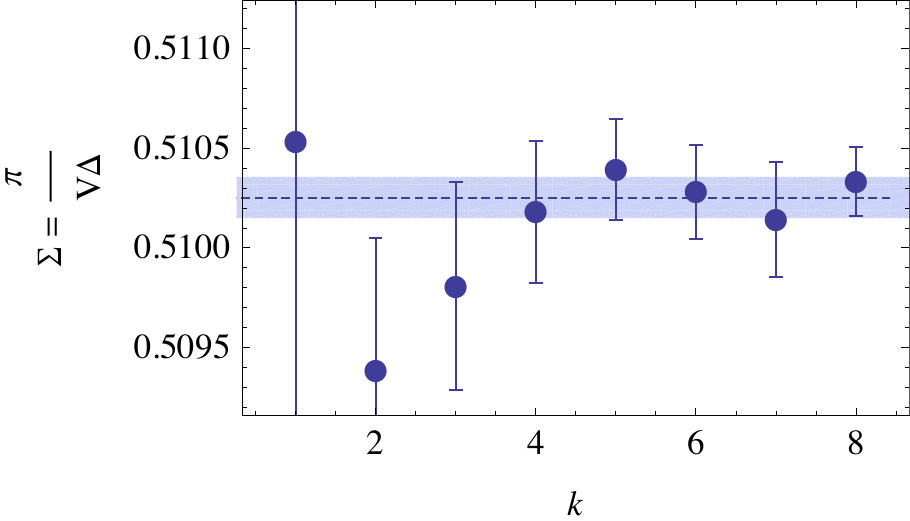}
\caption{
Eight smallest EV distributions
of chGUE ($\beta=2$),\ $n=\nu=0$
fitted to the Dirac spectrum of compact U(1) LGT [left].
Combining 8 values of the mean level spacing $\Delta$
obtained by fitting each distribution, 
the quenched chiral condensate $\Sigma={\pi}/{V\Delta}$ is determined 
as precisely as $\Sigma a^3=0.51025(10)$.
This precision cannot be hoped for from fitting the spectral density
or any single individual EV distribution [right].}
\end{center}
\end{figure}

Among established techniques of computing so-called gap probabilities,
neither the construction of Fredholm eigenfunctions/eigenvalues a la Gaudin-Mehta \cite{Gau61},
of linear differential equations a la Edelman \cite{Ede88},
or of Painlev\'{e} transcendental equations a la Tracy-Widom \cite{TW94} (in historical order)
is promising for 
the individual distribution of the $k^{\mathrm{th}}$ smallest eigenvalue
of the massive chiral ensemble (\ref{jpdchiralmassiv}).
Extending Forrester-Hughes' work \cite{FH94}, 
we thus proposed an alternative method \cite{DN01} consisting of 3 steps:
(i) relate the joint distribution of the first $k$ eigenvalues,
$\frac{1}{(N-k)!}\int_{x_k}^\infty dx_{k+1}\cdots
\int_{x_k}^\infty 
dx_N\,P(\{x\})$,
to the normalization (partition function) $\mathcal{Z}$ with $\beta k+\beta(\nu+1)/2-1$ additional masses and
a fixed topological charge $2/\beta+1$, by a shift of all variables by $x_k$,
(ii) replace $\mathcal{Z}$'s by their microscopically-scaled forms \cite{GW96,NN00}
as functions of scaled variables
$\zeta_i=\sqrt{8N x_i}$ and $\mu_f=\sqrt{8N}m_f$,
and (iii) integrate over the 
variables
$\zeta_1, \ldots, \zeta_{k-1}$ in a cell
$0\leq \zeta_1 \leq \cdots \leq \zeta_{k-1}\leq \zeta_k$.
This simple procedure leads to compact expressions for 
some easy cases, such as $\beta=2, n=0, k=1$ for which
$p_1(\zeta)=(\zeta/2) \e^{-\zeta^2/4}\det[I_{f-g+2}(\zeta)]_{f,g=1}^\nu$ \cite{FH94}.
However, the first and second steps entail
restrictions to massive Dirac fermions and
on the parity of the number of massless fermions ($\beta=4$) or 
of the topological charge ($\beta=1$),
and the $(k-1)$-fold numerical integration in the third step becomes
exponentially resource-consuming as $k$ increases\footnote{%
$p_{1\sim6}(\zeta)$ for $\beta=2$, $n=1, 2$ calculated in this way
excellently fit the data from dynamical QCD simulations \cite{Dam00}.}.
In this report we show that
these technical problems can be partly circumvented
by utilizing the quadrature method applied to Fredholm determinants \cite{Bor10}.

\section{Individual eigenvalue distribution}
Consider a stochastic distribution of $N$ points $\{x_1, \ldots, x_N\}$ on a real line.
Let $E_k(J)$ denote the probability 
of finding exactly $k$ points from 
in a (joint of) interval(s) $J$.
Then their generating function $E({z};J)$
is related to the $p$-point correlation functions of the densities
$R_p(x_1,\ldots,x_p):=\left\langle\prod_{k=1}^p \rho(x_k)\right\rangle$, $\rho(x):=\sum_{i=1}^N \delta(x-x_i)$,
by
\be
E({z};J):=\sum_{k\geq 0} (1-{z})^k E_k(J)
=1+
\sum_{p\geq 1}
\frac{(-{z})^p}{p!}
\int_J dx_1 \cdots\int_J dx_p\,R_p(x_1,\ldots,x_p).
\label{ER}
\ee
Eigenvalues of a unitary ensemble ($\beta=2$) obey a determinantal point process,
i.~e.\ their $p$-point correlation function is written as a $p\times p$ determinant
of a scalar kernel $K(x,x')$
constructed from orthonormal functions $\psi_i(x)$ corresponding to the weight
in (\ref{chGbE}),
\be
R_p(x_1,\ldots,x_p)=\det [K(x_i, x_j)]_{i,j=1}^p ,
\ \ \ 
K(x,x'):=\sum_{i=1}^N \psi_i(x)\psi_i(x').
\label{DPP}
\ee
Likewise, eigenvalues of an orthogonal ensemble ($\beta=1$),
a symplectic ensemble ($\beta=4$), or ensembles interpolating two different classes
(Dyson's Brownian motion model)
obey a Pfaffian point process,
i.~e.\ their $p$-level correlation function is written 
as a $p\times p$ quaternion determinant of a quaternionic kernel $\bs{K}(x, x')$
or a $2p\times 2p$ Pfaffian of its $\mathbb{C}$-number $2\times 2$-matrix representative
(denoted by the same $\bs{K}(x, x')$ for notational simplicity),
\be
R_p(x_1,\ldots,x_p)=\qdet [\bs{K}(x_i, x_j)]_{i,j=1}^p =
{\rm Pf}\, Z[\bs{K}(x_i, x_j)]_{i,j=1}^p,
\label{PPP}
\ee
($Z=i\sigma_2\oplus\cdots\oplus i\sigma_2$ denotes a skew-unit matrix),
constructed from skew-orthonormal functions.
Due to these determinantal properties (\ref{DPP}), (\ref{PPP})
and the identity (\ref{ER}), 
the generating function
$E({z};J)$ 
is expressed as a Fredholm (quaternion-)determinant
\be
E({z};J)={\rm det} ({\mathbb I}-{z} \hat{K}_J)\ \mbox{(scalar)}\ \ \mbox{or}\ \ 
{\rm qdet} ({\mathbb I}-{z} \hat{\bs{K}}_J)\ \mbox{(quaternion)},
\label{FredholmDet}
\ee
where $\hat{K}_J$ and $\hat{\bs{K}}_J$ denote integral operators acting on the Hilbert spaces of
1- and 2-component
$\mathcal{L}^2$ functions on $J$ with involution kernels $K(x,x')$ and $\bs{K}(x,x')$, respectively.
From (\ref{ER}), (\ref{FredholmDet}), and the relationship between a Pfaffian 
and a  determinant 
${\rm Pf}={\rm det}^{1/2}$,  $E_k(J)$ is expressed as
\be
E_k(J)=
\frac{1}{k!} \(-\partial_{z}\)^k \left. {\rm det} ({\mathbb I}-{z} \hat{K}_J)\right|_{{z}=1}\ \mbox{(scalar)}\ \ \mbox{or}\ \ 
\frac{1}{k!} \(-\partial_{z}\)^k \left.{\rm det} ({\mathbb I}-{z} \hat{\bs{K}}_J)^{1/2}\right|_{{z}=1}\ \mbox{(quaternion)}.
\label{EkFredholmDet}
\ee
Now we set $J=[0,s]$ and abbreviate $\hat{K}_s:=\hat{K}_{[0,s]}$, $E_k(s):=E_k([0,s])$, etc.
By performing ${z}$-derivatives explicitly\footnote{%
An alternative method is to compute ${\rm det} (1-{z} \hat{K}_J)$ 
for complex $z$'s on a circle centered at $z=1$ and evaluate the Cauchy integral 
$\oint dz\, (z-1)^{-k-1} {\rm det} (1-{z} \hat{K}_J)$ numerically \cite{Bor09}.
Our method of computing $T_1, \ldots, T_k$ in required precision is more suited for $k\lesssim 10$;
the Cauchy method becomes advantageous for $k\gtrsim 10$,
as the increased cost (by the number of points discretizing the circle)
is compensated by the convenience of computing $E_0, \ldots, E_{k-1}$ all at once.
}, $E_k(s)$ are expressed by Fredholm determinants at ${z}=1$
and the functional traces of the resolvents
$T_n(s)=
{\rm tr} \bigl(\hat{{K}}_s({\mathbb I}-\hat{{K}}_s)^{-1}\bigr)^n
\ \mbox{(scalar)}$ or 
${\rm tr} \bigl(\hat{\bs{K}}_s({\mathbb I}-\hat{\bs{K}}_s)^{-1}\bigr)^n/2\ \mbox{(quaternion)}
$
as
\begin{eqnarray}
\!\!\!\!\!\!&&
E_0(s)=
\det({\mathbb I}-\hat{K}_s) \ \ \mbox{(scalar)}\ \ \mbox{or}\ \ 
\det ({\mathbb I}-\hat{\bs{K}}_s)^{1/2}\ \  \mbox{(quaternion)},
\label{Ek}
\\
\!\!\!\!\!\!&&
E_1(s)=E_0(s) T_1, \quad
E_2(s)= \frac{E_0(s)}{2!} \(T_1^2 -T_2 \),\quad
E_3(s)= \frac{E_0(s)}{3!} \(T_1^3 -3T_1 T_2  +2T_3 \),
\nn\\
\!\!\!\!\!\!&&
E_4(s)= \frac{E_0(s)}{4!} \(T_1^4 -6 T_1^2 T_2 + 3T_2^2 +8 T_1 T_3 -6 T_4 \) ,
\nn\\
\!\!\!\!\!\!&&
E_5(s)= \frac{E_0(s)}{5!} \(T_1^5 - 10 T_1^3 T_2 + 20 T_1^2 T_3 + 15 T_1 T_2^2 
 - 30 T_1 T_4 - 20 T_2 T_3 + 24 T_5\),
\nn\\
\!\!\!\!\!\!&&E_6(s)=\frac{E_0(s) }{6!}
\left\{
{{T_1^6 - 15T_1^4 T_2 + 40T_1^3 T_3 + 45T_1^2 T_2^2 - 90T_1^2 T_4 - 120T_1 T_2 T_3}
      \atop
{- 15T_2^3 + 144T_1 T_5 + 90T_2  T_4 + 40T_3^2 -120 T_6}}
\right\},
\nn\\
\!\!\!\!\!\!&&E_7(s)=
\frac{E_0(s) }{7!}
\left\{
{{T_1^7 
- 21T_1^5 T_2  
+ 70T_1^4 T_3 
+ 105T_1^3 T_2^2 
- 210 T_1^3 T_4 
- 420T_1^2 T_2  T_3
- 105 T_1 T_2^3
+ 504 T_1^2  }
\atop
{
\times T_5 + 630 T_1 T_2 T_4
 + 280 T_1 T_3^2 
 +   210 T_2^2 T_3 
 - 840 T_1 T_6
 -  504 T_2 T_5
 - 420 T_3 T_4
 + 720 T_7}}
\right\} .
\nn 
\end{eqnarray}
Then the individual  distribution $p_k(s)$ of the $k^{\rm th}$ smallest positive eigenvalue
is expressed in terms of
$E_0(s), \ldots , E_{k-1}(s)$ as
$p_k(s)=-\partial_s \sum_{\ell=0}^{k-1} E_{\ell}(s).$

An efficient way of numerically evaluating the Fredholm determinant of
a trace-class operator $\hat{{K}}_s$ or $\hat{\bs{K}}_s$
is the Nystr\"{o}m-type discretization \cite{Bor10}
\be
{\rm det}({\mathbb I}-\hat{{K}}_s)\simeq \det
\left[
\delta_{ij}-K(x_i,x_j) \sqrt{w_i\,w_j}\right]_{i,j=1}^m,
\ \ 
{\rm det}({\mathbb I}-\hat{\bs{K}}_s)\simeq \det
\left[
I\delta_{ij}-\bs{K}(x_i,x_j) \sqrt{w_i\,w_j}\right]_{i,j=1}^m .
\label{Nystrom}
\ee
Here we employ a quadrature rule consisting of a set of points $\{x_i\}$ 
taken from the interval $[0,s]$ and associated weights $\{w_i\}$ such that
${\int_0^s f(x)dx  \simeq \sum_{i=1}^m f(x_i) w_i}$.
Similarly the functional trace in
$T_n(s)$ can be approximated by its discretized version.
For a practical purpose we choose the Gauss-Legendre quadrature rule, i.e.\ sampling $\{x_i\}$ from
the nodes of Legendre polynomials normalized to $[0,s]$,
so that the discretization gives an exact value for an arbitrary $(m-1)^{\rm th}$-order polynomial $f(x)$. 
Exponential convergence in $m$ to the exact value of the Fredholm determinant is guaranteed
for any smooth kernel.
This method was previously applied to the quenched ($n=0$) chiral ensemble
interpolating chGSE and chGUE,
which enabled precise determination of the
pion decay constant in the SU(2) gauge theory \cite{NY14}.
There we found it far more than sufficient to set the approximation order $m$ to be $\lesssim100$
for the purpose of fitting the lattice Dirac spectra.

\section{Chiral ensembles}
The scalar kernel describing the eigenvalue correlation of the chiral GUE
($\beta=2$)
in the microscopic scaling limit is given in Eq.(33) of Ref.\cite{DN98}.
As an illustrative example that is applicable for QCD with light flavors,
we exhibit $p_{1\sim 8}(\zeta)$ with $n=2+1$ (Fig.~2),
computed by Nystr\"{o}m-type discretization of the Fredholm determinant and traces (\ref{Ek}) of
the kernel in the confluent limit,
\bea
&&
K(\zeta,\zeta';\mu, \mu, \mu')=
\frac{\sqrt{\zeta \zeta'}}{(\zeta'{}^2-\zeta^2)
(\zeta^2+\mu^2)(\zeta'{}^2+\mu^2)\sqrt{(\zeta^2+\mu'{}^2)(\zeta'{}^2+\mu'{}^2)}}
\frac{\det{B}(\zeta,\zeta';\{\mu\})}{\det{A}(\{\mu\})},
\nonumber\\
&&
A=\left[
\begin{array}{c}
(-\mu)^{g-1} I_{\nu+g-1}(\mu)\\
(-\mu)^{g-2} I_{\nu+g-2}(\mu)\\
(-\mu')^{g-1} I_{\nu+g-1}(\mu')
\end{array}
\right]_{g=1,2,3},\ \ 
B=
{\small \left[
\begin{array}{c}
\zeta^{g-1} J_{\nu+g-1}(\zeta)\\
\zeta'{}^{g-1} J_{\nu+g-1}(\zeta')\\
(-\mu)^{g-1} I_{\nu+g-1}(\mu)\\
(-\mu)^{g-2} I_{\nu+g-2}(\mu)\\
(-\mu')^{g-1} I_{\nu+g-1}(\mu')
\end{array}
\right]_{g=1,\ldots,5} .}
\label{K2_2+1}
\eea
\begin{figure}[b]
\begin{center}
\includegraphics[bb=0 0 360 238,width=49mm]{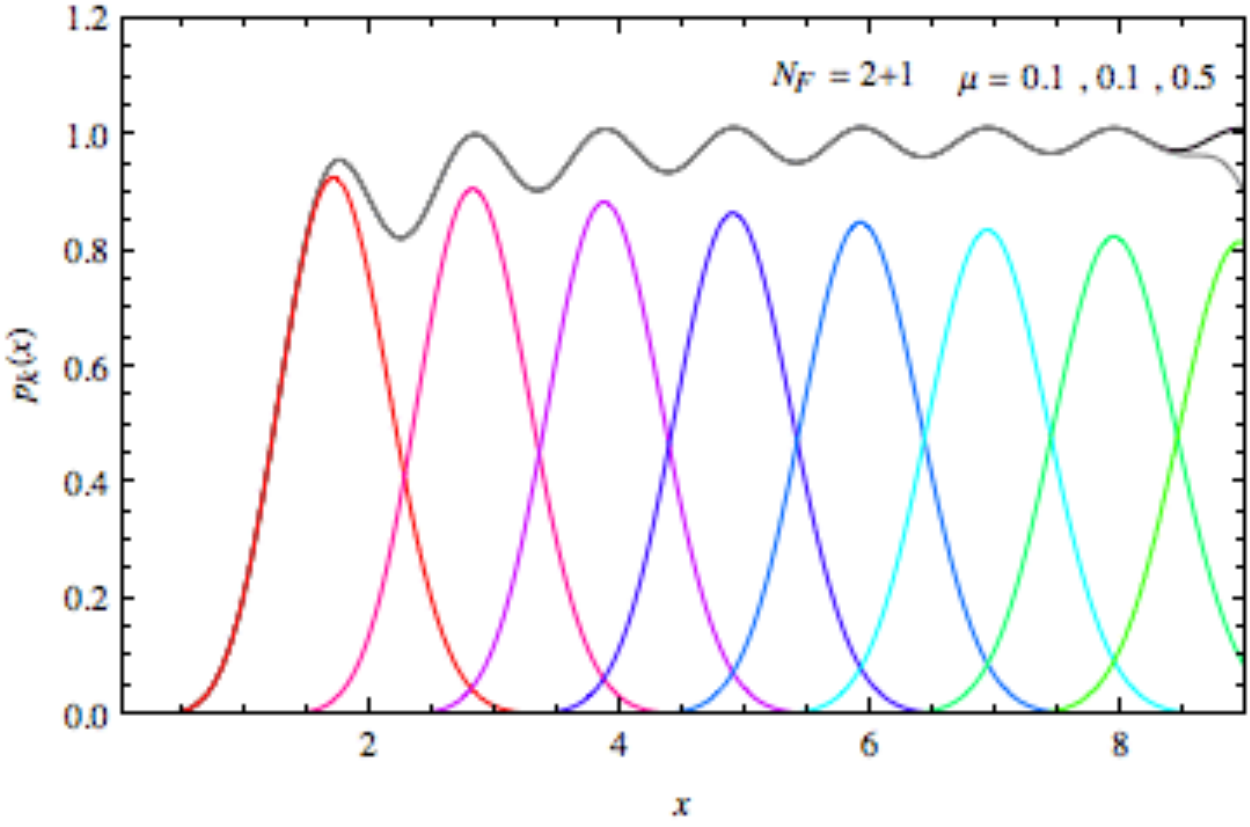}~
\includegraphics[bb=0 0 360 238,width=49mm]{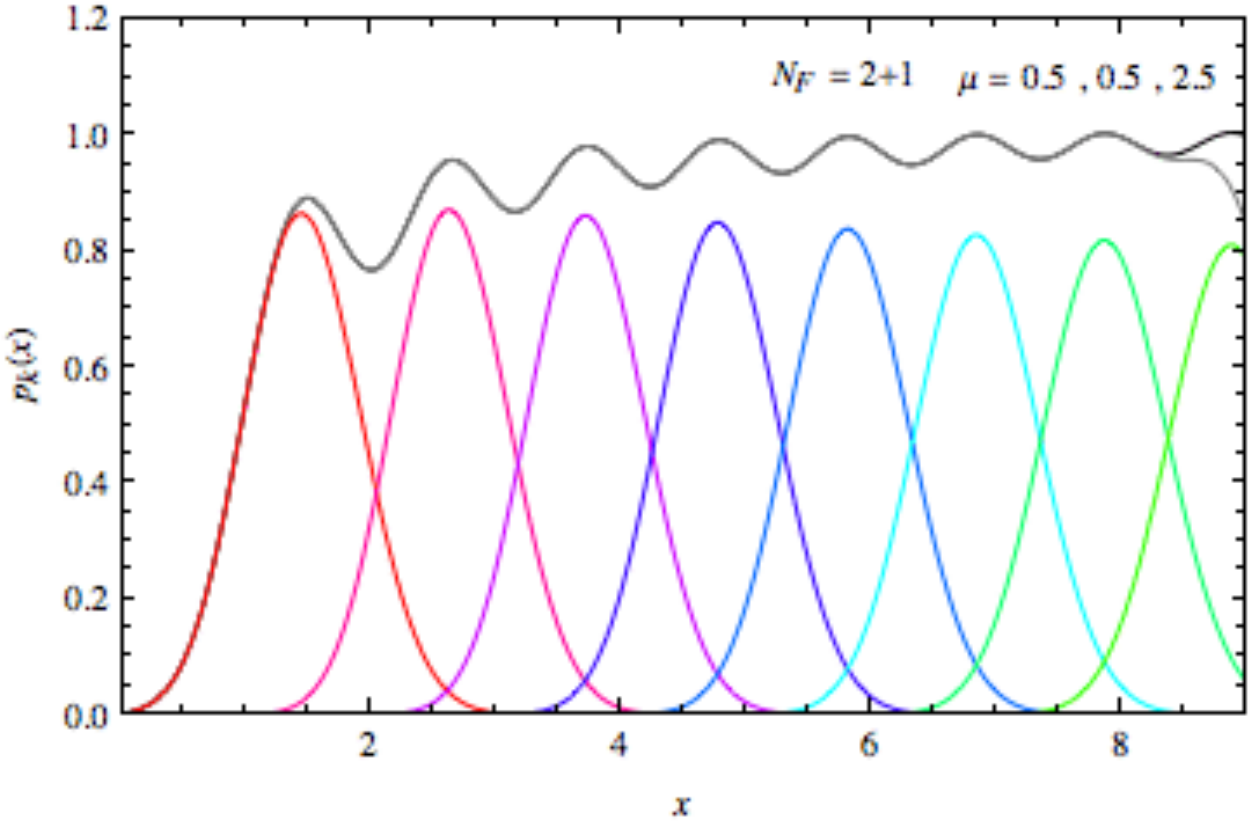}~
\includegraphics[bb=0 0 360 238,width=49mm]{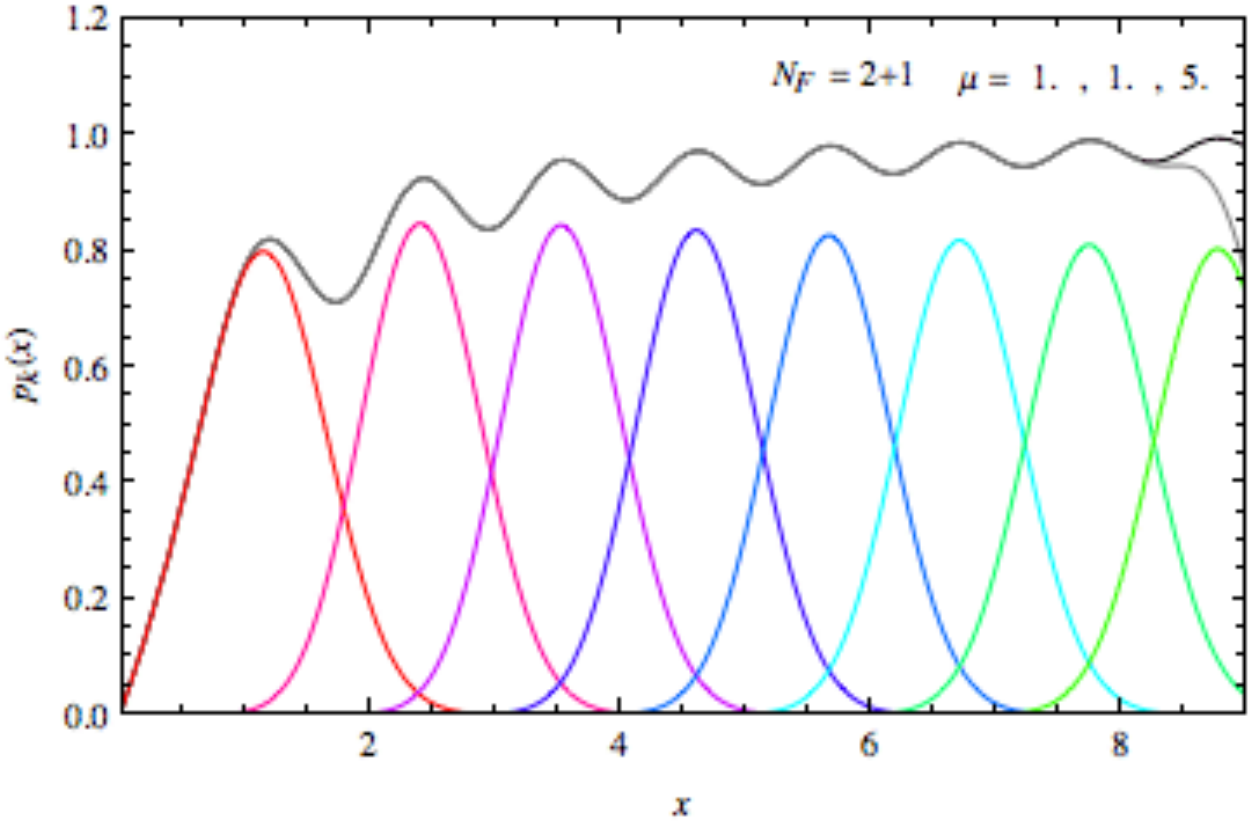}\\
\includegraphics[bb=0 0 360 238,width=49mm]{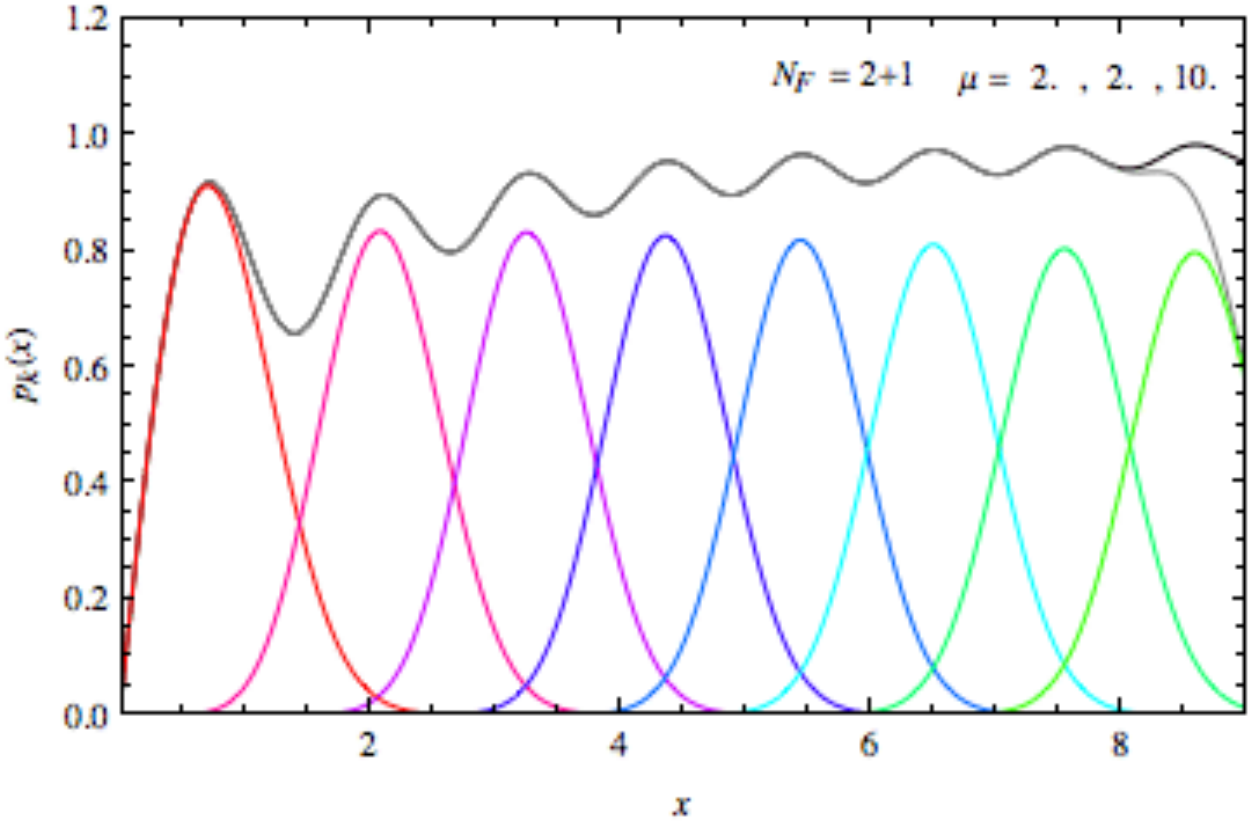}~
\includegraphics[bb=0 0 360 238,width=49mm]{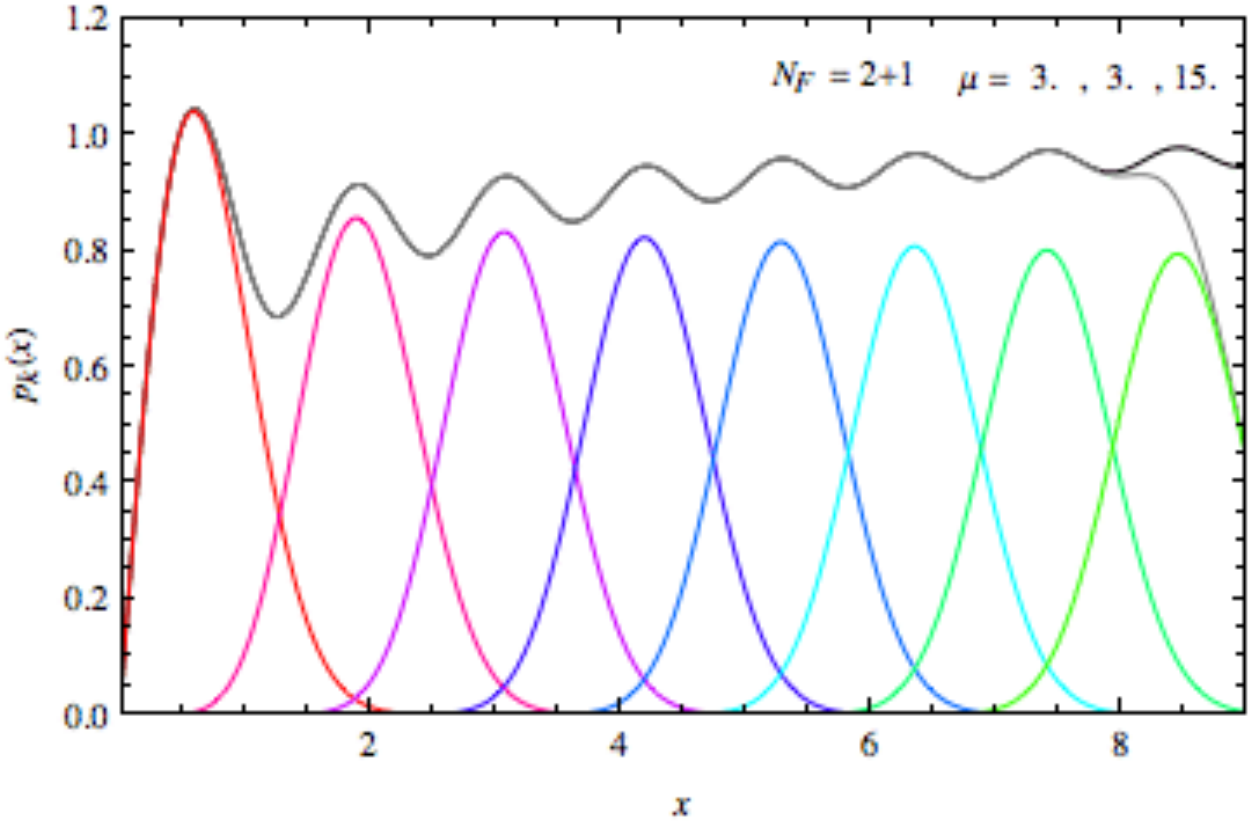}~
\includegraphics[bb=0 0 360 238,width=49mm]{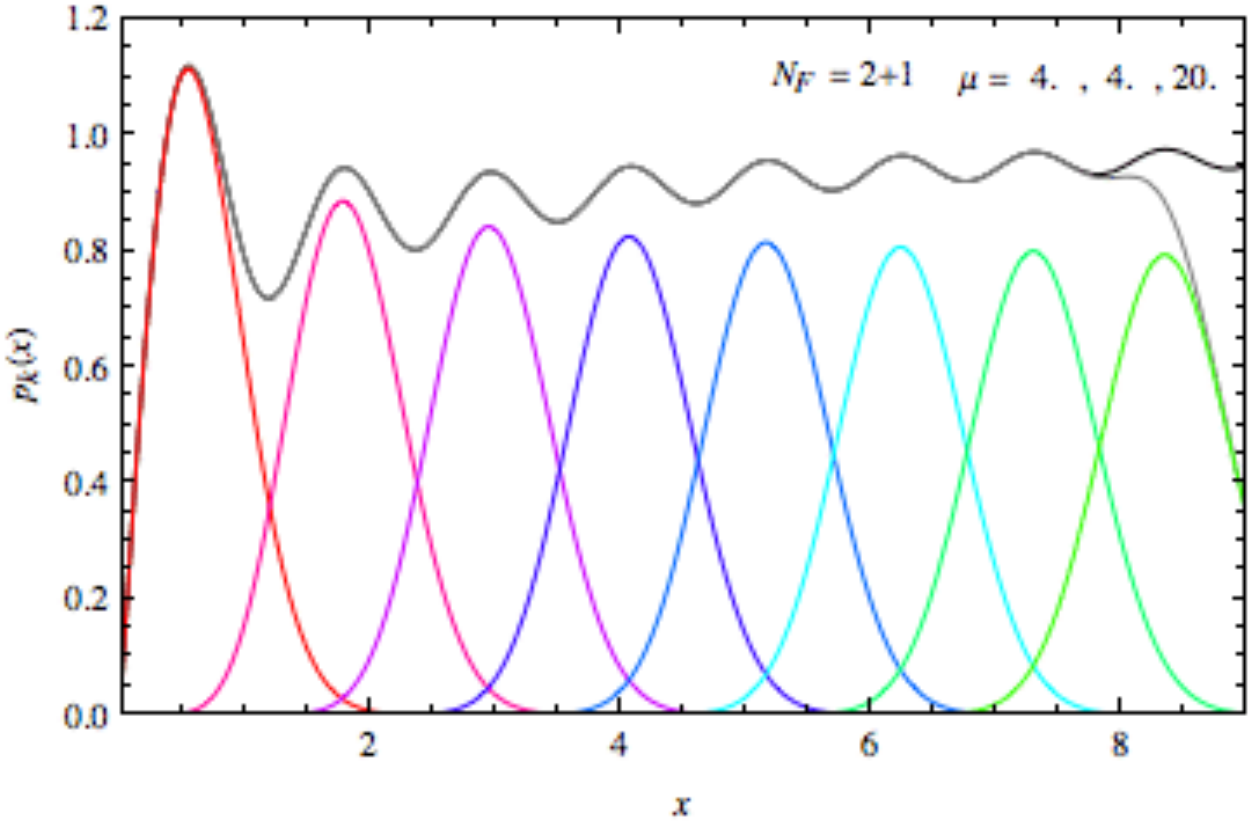}
\caption{
8 smallest EV distributions
of massive chGUE, $n=2+1, \nu=0$,
with $\pi\mu =\pi\mu'/5=0.1\sim 4.0 $.}
\end{center}
\end{figure}
Note that the scalar kernel (\ref{DPP}) in the scaling limit is a self-convolution of functions\footnote{%
Normalization of the scaled orthogonal polynomials at generic $t\in [0,1]$ was not explicitly given in Ref.\cite{DN98}.}
$V(t,\zeta)$
which
originate from
the orthonormal functions $\psi_i(x)$ with $t=i/N$ and $\zeta=\sqrt{8N x}$ ($\{\mu\}$ is suppressed),
\bea
&&K(\zeta, \zeta')=\int_0^1 dt\,V(t,\zeta)V(t,\zeta')=({V}^T\circ {V})(\zeta,\zeta'),
\ \ \ \ \int_0^\infty d\zeta\,V(t,\zeta)V(t',\zeta)=\delta(t-t')
\nn\\
&&
V(t,\zeta)=
\frac{\sqrt{\zeta/2}}{\prod_{f=1}^n \sqrt{\zeta^2+\mu^2_f}}
\frac{\det{\tilde{B}}(t; \zeta)}{\det{\tilde{A}}(t)},
\label{V}\\
&&
\tilde{A}=\left[(-\mu_f)^{g-1} I_{\nu+g-1}(\sqrt{t}\mu_f)\right]_{f, g=1}^n,\ \ 
\tilde{B}=\left[
\begin{array}{c}
\zeta^{g-1} J_{\nu+g-1}(\sqrt{t}\zeta)\\
\left[(-\mu_f)^{g-1} I_{\nu+g-1}(\sqrt{t}\mu_f)\right]_{f=1}^n
\end{array}
\right]_{g=1}^{n+1} .
\nn
\eea
Thus its Fredholm determinant factorizes into the product of two Fredholm determinants \cite{For06},
\be
E(z;[0,s])=
{\rm det}({\mathbb I}+\sqrt{z} \hat{V}_s){\rm det}({\mathbb I}-\sqrt{z} \hat{V}_s),
\label{VV}
\ee
where $\hat{V}_s$ is an integral operator $\mathcal{L}^2[0,1]\to \mathcal{L}^2[0,1]$
(the domain of definition is rescaled from $[0,s]$)
of convolution by a rescaled kernel $\sqrt{s} V(t, s \zeta)$.
Numerical evaluation of (\ref{VV}) gives an alternative, convenient way of computing $p_k(s)$
of chiral GUE.

In the case of chiral GSE ($\beta=4$),
the partition function and correlation functions, from which
the quaternion kernel can easily read off,
have readily been worked out in Refs.~\cite{NN00}.
Thus one can simply apply the quaternion side of (\ref{Nystrom}) to compute $p_k(\zeta)$.
However, we must admit that
the case of chiral GOE ($\beta=1$) at even $\nu$, which lay outside of the scope of Ref.\cite{DN01},
still poses a problem; it is
because the exponential convergence of the quadrature approximation is not
guaranteed for its quaternion kernel which
includes a discontinuous signature function.

\section{q-Hermite ensemble}
Among the random matrix ensembles that possess weak confining potentials ($\sim \tr (\log H)^2$) and
break the Wigner-Dyson universality,
the so-called $q^{-1}$-Hermite ensemble \cite{IM94}
associated with the Askey weight 
leads to a translationally invariant
spectrum in the bulk and  serves as a candidate of a phenomenological model
describing the mobility edge of disordered Hamiltonians.
As $q$ is deformed from unity,
its kernel in far away from the origin,
\be
K_q(\zeta, \zeta')=
\frac{a \vartheta _1(\pi  (\zeta-\zeta'),q)}{\pi  \vartheta _1^{\prime }(0,q) \sinh(a (\zeta-\zeta')) },
\ \ 
q:=\e^{-{\pi^2}/{a}},
\label{ellip}
\ee
interpolates between Wigner and Poisson statistics
without crystallization of eigenvalues as seen in other ensembles \cite{MCI01}.
Here we slightly extend our previous observation that by tuning a single parameter $q$, the 
level spacing distribution of this deformed ensemble can perfectly fit the 
scale-invariant critical statistics of the mobility edge of Anderson Hamiltonians \cite{Nis99},
by considering the distribution of spacings between
$k^{\rm th}$ nearest-neighboring levels, $p_k(s)$. 
This quantity is related to the gap probabilities $E_\ell([-s/2,s/2])$ as
$p_k(s)=\partial_s^2\sum_{\ell=0}^{k-1}E_\ell([-s/2,s/2])$, which we evaluate by the quadrature approximation
of the Fredholm determinant and traces of the elliptic kernel (\ref{ellip}).

In Fig.3 we performed one-parameter fitting of $p_k(s)$ to the spacing data from the mobility edges
(their locations are known from previous studies) of
the Anderson Hamiltonian in the unitary class
and the stagger Dirac operator of high-temperature QCD \cite{Nis13}.
Impressive fittings especially in the case of Anderson Hamiltonian again recalls us the power of
combining the information from multiple level spacing distributions;
if applied to the QCD data on larger lattices, it should help identifying the location of the
mobility edge and the critical parameter in high precision.

\begin{figure}[tbh]
\begin{center}
\includegraphics[bb=0 0 344 222,width=75mm]{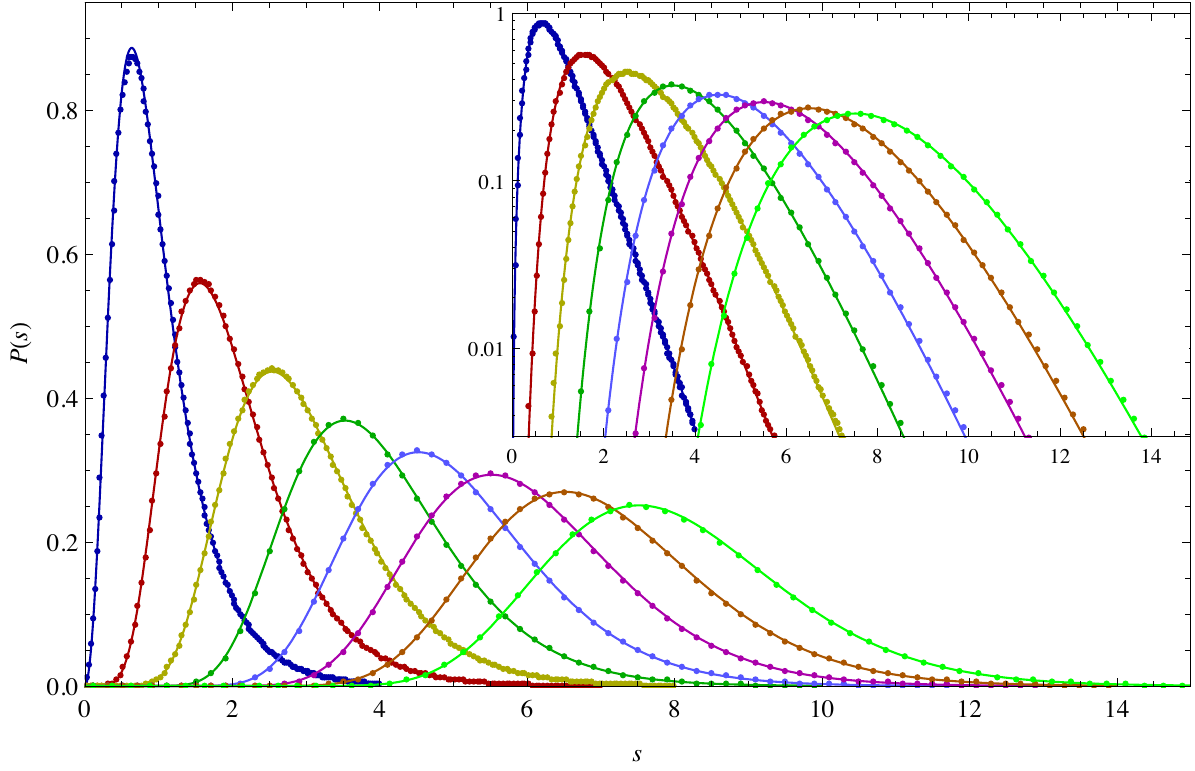}~
\includegraphics[bb=0 0 353 247,width=75mm]{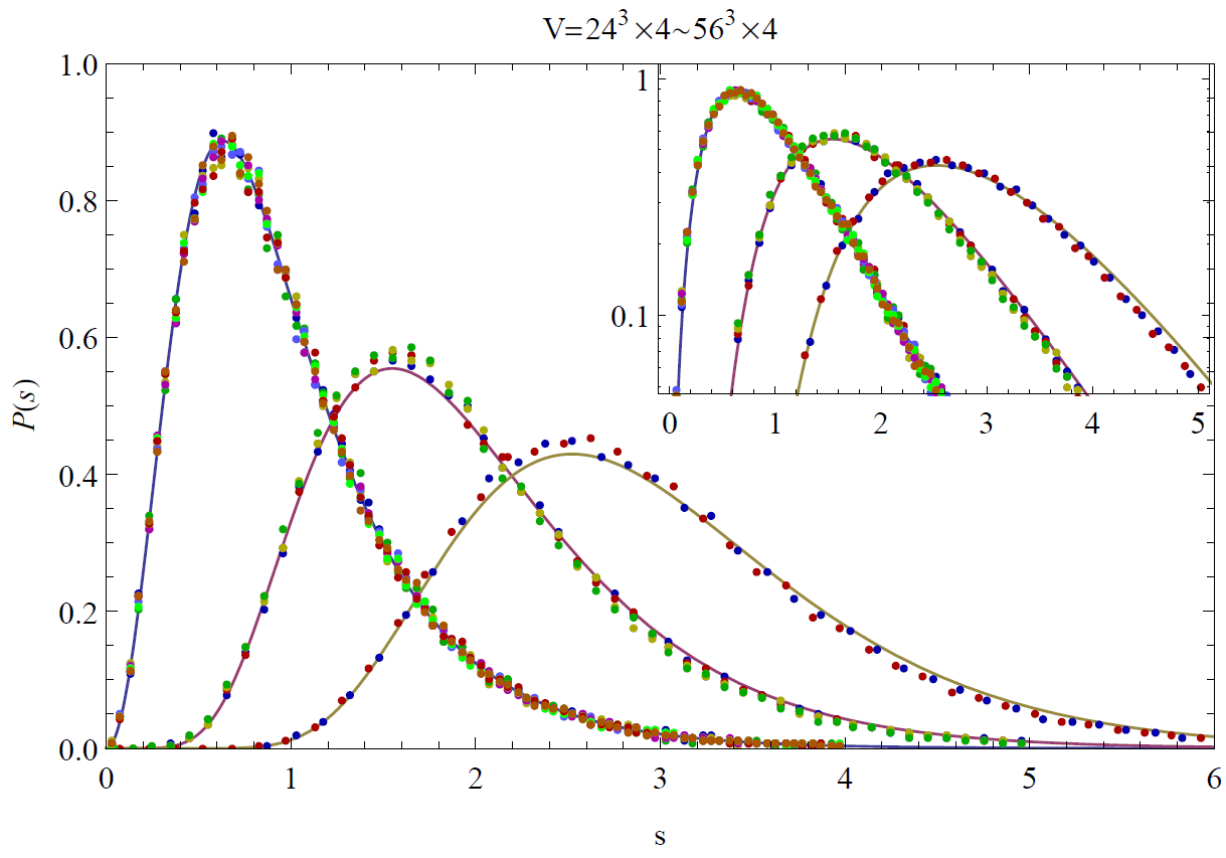}
\caption{
Distributions of spacings between $k^{\rm th}$ nearest-neighboring levels $p_k(s)$, $k=1,\ldots, 8$
of the $q^{-1}$-Hermite ensemble
fitted to the mobility edges of
unitary Anderson Hamiltonian on $V=20^3$ [left]
and of high-temperature QCD Dirac operator $(T=2.6T_c)$ on $V=(24^3\sim 56^3)\times 4$
(eigenvalue data provided by T. Kov\'{a}cs) [right].
Parameters and lattice settings are the same as in Figs.2 and 6 of Ref.\cite{Nis13}.}
\end{center}
\end{figure}

\acknowledgments
I thank Simons Center for Geometry and Physics for 
invitation to the program
``Foundations and Applications of 
Random Matrix Theory 
in Mathematics and Physics"
during which this report was partly written,
its participants,
and especially the organizers J.~Verbaarschot and P.~Forrester for illuminating discussions and hospitality.
I also thank T. Kov\'{a}cs for kindly providing the lattice Dirac spectral data presented in Fig.3 [right],
whose eigenvectors are analyzed in Ref.\cite{Kov15}.

\end{document}